\documentstyle[12pt]{article}
\def\l{\Lambda}
\def\al{\bar\Lambda}

\begin{document}
\begin{titlepage}
\begin{center}
 {\Large\bf Strangeness production in nuclear interactions
at 200AGeV and the number of nucleon- nucleon collisions}
 \end{center}
\vspace{1cm}
\begin{center}
{\large Roman Lietava${}^{a),b)}$ and J\'an Pi\v s\'ut${}^{a),c)}$}
\end{center}
\vspace{1cm}
 \begin{center}
{\it  
${}^{a)}$Department of Physics,Comenius University, \\
      SK-842 15 Bratislava,Slovakia \\
${}^{b)}$GRPHE,Universite de Haute Alsace,Mulhouse,France\\
${}^{c)}$Laboratoire de Physique Corpusculaire,\\
Univ. Blaise Pascal, F-63177 Aubiere, France

     }
\end{center}
\vspace{1cm}
\abstract

{ Data on mean numbers of  
$\Lambda$,  $\bar\Lambda$, K's and on the
 total number $<s\bar s>$ of pairs of
strange valence quarks in final state hadrons in hadronic and nuclear
collisions at CERN- SPS energies are studied as a function
of  the mean number 
$<n_{coll}>$ of nucleon- nucleon collisions. Results give indications of 
an almost linear dependence over most of the region of 
$<n_{coll}>$. This in turn
points out to strangeness being produced mostly in the central rapidity
region of nucleon- nucleon collisions by a mechanism similar to a hard or
semi- hard process. The available data are extrapolated to Pb+Pb interactions
by two simple models, leading to $<s\bar s>_{Pb+Pb}\approx 300\pm 30 $.
 
 Observations exceeding these values of
$<s\bar s>_{Pb+Pb}$ would give evidence of the onset of a
 new dynamical regime in Pb+Pb interactions.}

\end{titlepage}
\section{Introduction}
  Production of strange particles in proton- proton (pp),
proton- nucleus (pA) and nucleus- nucleus (AB)
collisions  has been extensively studied both experimentally and 
theoretically in the past decade. 

The enhanced production of strange particles especially of antibaryons
 has  been suggested as one of possible signatures of Quark-Gluon
Plasma (QGP) formation see e.g. Refs.\cite{r1}. Alternative scenarios 
for strangeness enhancement are hadron gas 
( for review see \cite{r2,r2.1} ) and microscopic models ( for reviews see 
\cite{r3,r3.1,r4,r5} ).

Ratios of mean numbers of $\Lambda$ and $K^0_s$ per event 
over negative hadrons
$h^-$, denoted as $<\Lambda>/<h^->$ and $<K^0_s>/<h^->$ has been measured
by NA-35  Collaboration \cite{r6,r7} and found to be 
about twice as large as the 
same
ratios in pA collisions \cite{r6,r7,r8,r9}.
  Compilations of data can be found in Refs.\cite{r10,r11,r12,r13}.

An alternative, and in a sense global, point of view on strangeness
enhancement consists in counting the total mean number $<s\bar s>$ of
s- and $\bar s$- valence quark pairs in final state hadrons
and studying the ratio \cite{r13,r14,r15,r16}

$$\lambda _s= \frac {<s\bar s>}{0.5(<d\bar d> +<u\bar u>)}$$

The value of $<s\bar s>$ pairs is experimentally measurable and given
as \cite{r13,r15}:

$$<s\bar s>\approx0.5(<\Lambda>+<\Sigma>+<\bar\Lambda>+<\bar\Sigma>$$

$$ + <K^0> + <\bar K^0>+<K^+> +<K^->)
 + 2/3<\eta>  $$

Analyzing the data on strangeness production in pp, pA and central S+S
collisions Bialkowska et al. \cite{r13} have found that for central S+S
interactions the value of $\lambda _s$ is about twice as large as for 
pA interactions. For pA interactions (see Table 3. in \cite{r13})
$\lambda_s \approx 0.2$, whereas for central S+S interactions
$\lambda_s = 0.36\pm0.03$

The dynamical mechanism responsible for hadron - and in particular for
strange hadron production - is at present unknown and it is therefore
difficult to see whether the increase of ratios  $<\Lambda>/<h^->$ and
$<K>/<h^->$ and of $\lambda_s$ when going from pA to central S+S
interactions is due to the formation of a new form of matter or whether
it is due to more mundane reasons.

In order to obtain some hint into possible reasons of the observed
strangeness enhancement we shall study in the next Sect. the dependence
of $<\Lambda>$, $<\bar \Lambda>$, $<K>$ and $<s\bar s>$ on 
the mean number of nucleon- nucleon
collisions $<n_{coll}>$ in pA and AB interactions.

The purpose of the present note is to show 
that $<\Lambda>,<K>$ and in particular
$<s\bar s>$ depend  linearly - with a mild attenuation increasing with
the number of collisions -
 upon $<n_{coll}>$ what in turn 
indicates that strangeness is produced by a mechanism reminiscent
of hard or semi- hard processes.

In Sect.3 we shall present two simple, and basically rather similar,
models describing strangeness production as a sum of contributions of
individual nucleon- nucleon collisions. The models permit
 to extrapolate the available data to the case of Pb+Pb interactions
 and the difference between the two models gives a feeling of the
 error of extrapolation.

If data on $<s\bar s>$ in Pb+Pb interactions were substantially
 higher than these simple
 extrapolations we would have an evidence of an onset of a new
dynamical regime between S+(heavy target) and Pb+Pb collisions.
 
The attenuated linear dependence of $<s\bar s>$, $<\Lambda>$,
 $<\bar \Lambda>$and $<K>$ on $<n_{coll}>$
is rather weird and we shall discuss a possible message of this
phenomenon in Sect.4.

\section{Dependence of $<\Lambda>,<\bar\Lambda>$,$<K^0_s>$ and
  $<s\bar s>$ on the number of nucleon- nucleon collisions}
\label{dep}
In Fig.1 we plot $<\Lambda>,<\bar\Lambda>$ and $<K^0_s>$ vs. the mean
number of nucleon- nucleon collisions $<n_{coll}>$ for pp, pA and AB
interactions. Values of $<n_{coll}>$ have been calculated by a simple
Monte Carlo model of the Glauber type to be described in more details in
the next Section.

  The data were compiled  from the original papers. The exact
references are quoted in Tab.1. 
It should be noted here that data of three experiments \cite{r7,r8,r9}
are not consistent. This can be seen 
by comparing kaons in pAg from \cite{r9} with
 kaons in pXe in \cite{r8}, kaons in pAr in \cite{r8} and kaons
in pS \cite{r7}.
 Quoted number of kaons in the same experiment
has also changed with increased statistics by more then 30\% \cite{r6,r7}.
We take the value from the more recent publication.

The value of $<s\bar s>$ pairs in pp collisions
is experimentally measurable and given as
\cite{r13,r15}:
\begin{equation}
<s\bar s> = <s\bar s>_B + <s\bar s>_M 
\label{eq1}
\end{equation}

where $<s\bar s>_B$ and $<s\bar s>_M$
are contributions from baryons and mesons respectively.  

$$ <s\bar s>_B \approx 0.5( <\Lambda>+<\bar\Lambda>+
 <\Sigma> +<\bar\Sigma>) + <\Xi> +<\bar \Xi> $$

where $\Lambda$ stands for $\Lambda + \Sigma^0$ , $\Sigma$ stands
for $\Sigma^+ +\Sigma^-$ and $\Xi$ for $\Xi^0 + \Xi^-$ and correspondingly
for the antiparticles.
 
  Hyperon yields are not well known.
An estimate of $\Sigma$ hyperons  is
given by Wroblewski empirical formula \cite{r16} 
$<\Sigma> \approx(0.6\pm0.1) <\Lambda>$.

A rough  estimate of $\Xi$ hyperons
can be obtained by
extrapolating the ratio $\Xi^- /\Lambda $  from lower energy data
to 200 GeV/c and using the available information on $\Lambda $ 
production at 200 GeV/c. For the extrapolation 
to $p_{lab}=$ 200 GeV/c we use the value
of $\Xi^-/\Lambda $ ratio of 
$0.016\pm 0.011$ at $p_{lab}=$19GeV/c \cite{a37} 
and $0.025\pm0.007$
at $p_{lab}= $28GeV/c \cite{a38}. Extrapolation is made linearly
in $\sqrt{s}$ with the condition that the ratio should be always
non-negative. The result is $\Xi^-/\Lambda=0.0034\sqrt{s}$.
For $p_{lab}=200GeV/c$ we get $ \Xi^-/\Lambda=0.066$
Assuming that in pp collisions according to valence quark
content of protons $\Xi^0/\Xi^-\approx 2$, the total $\Xi$
contribution is
$<\Xi^-+\Xi^0>$ $\approx $ $3*\Xi^-/\Lambda <\Lambda>\approx
(0.2\pm0.05)<\Lambda>$.
In the $<\bar\Xi>$ sector the ratio $\bar\Xi^-/\bar\Lambda=0.06\pm0.02$
at $\sqrt{s}=63GeV$ was measured in Ref.\cite{a39}. For simplicity
we shall use $<\bar\Xi>=0.2<\bar\Lambda>$ which slightly overestimates
$<\bar\Xi>$ contribution to $<s\bar s>_B$.
The error on final $<s\bar s>_B$
is of the order of 1\%.

 For pp interaction at 200 GeV/c we thus have
$$<s\bar s>_B=(1.0\pm0.1)(<\Lambda>+<\bar\Lambda>)$$
and by using the compilation \cite{r12} this leads to
$<s\bar s>_B=0.109\pm0.015$.

The $<s\bar s>$ content of the meson sector can be approximated as
$$<s\bar s>_M\approx 0.5*(<K^0>+<\bar K^0>+
   <K^+>+<K^->) 
+{2\over 3}<\eta>
$$
where the $<\Phi>$ contribution is effectively included in
kaons and $\eta\prime$ is neglected.

 The $\eta$ production was measured
to be $0.20\pm0.02$ at $p_{lab}=400$ Gev/c \cite{b39}, 
$0.92\pm0.58$ at $\sqrt{s}=$53GeV/c \cite{b40} and less then 0.2
at $p_{lab}=$69GeV/c \cite{b41}.
Interpolating this linearly in $ln(s)$ to $p_{lab}=200GeV/c$
we get $<\eta>=0.18\pm0.05$

Summing up contributions of all mesons we get at $p_{lab}=200$ GeV/c\\
$<s\bar s>_M$ $=0.52\pm0.08$

Following 
Bialkowska et al. \cite{r13} we define
$$
R_{SB}={{<s\bar s>_B}\over {<\Lambda +\bar\Lambda>}}
$$
and
$$ R_{SM}= {<s\bar s>_M\over <K>}$$
where
$$ <K>=(2<K_s^0>+<K^+>+<K^->)/4$$

We estimate the total number of $s\bar s $ pairs produced in
AB (pA) collisions  as

\begin{equation}
<s\bar s>= R_{SB}(pp)<\Lambda+\bar\Lambda>_{AB}+R_{SM}(pp)<K>_{AB}
\end{equation}

Our values of parameters $R_{SM}$ and $R_{SB}$  are
$$
R_{SB}(pp)=1.0\pm 0.1 \qquad R_{SM}(pp)=2.6\pm 0.3
$$

whereas Bialkowska et al. \cite{r13} have found 

$$
R_{SB}(pp)=0.93\pm 0.06 \qquad R_{SM}(pp)=2.75\pm 0.09
$$
 
In our analysis of data on kaon production 
 in nuclear collisions whenever  experimental
data on charged kaons were not available the $<K>$ value was calculated
 assuming that the relative yields of $K_s^0$, $K^+$ and $K^-$
for a given AB (pA) reaction are the same as in nucleon-nucleon
(proton-nucleon) reaction.
We also take into account that for isospin zero system it holds
$  <K^+> +  <K^->\approx 2<K^0_s> $    

In Fig.2 we present the dependence of the mean number of $s\bar s$
pairs  on $<n_{coll}>$.

 Values of $<s\bar s>$ were calculated according to  Eq. (2)
The results agree with those of Bialkowska et al. \cite{r13} 
except for the pS value where we have taken more recent data \cite{r7}.

The Eq.(2) was derived assuming that $\Lambda$ and $\bar\Lambda$ are
corrected for feed down from week $\Xi$ and $\bar\Xi$ decays.
This is not the case for the NA35 data used in our analysis.
The effect of feed down can be estimated using the WA85 (SW collisions)
and WA94 data (SS collisions) \cite{wa85,wa94}. These data shows
that about 10\% of $\Lambda$ are from $\Xi$ and about 20\%
of $\bar\Lambda$ are from $\bar\Xi$. This will lower the $<s\bar s>$
values in SS and SAg collisions by about 4\%.

\section{ Extrapolations of strangeness production to Pb+Pb interactions}
\label{extrap}
In order to extrapolate strangeness production as described by $<s\bar s>$,
$<\Lambda>,<\bar\Lambda>$ and $<K^0_s>$
to the case of Pb+Pb collisions we shall first "parametrize" the data for
 collisions induced by lighter ions by a simple model.

Since the data on $<s\bar s>$ production are more stable with respect to
final state interactions like $\bar\Lambda N\to K\pi $ or
 $\bar KN\to \Lambda \pi $ - although not with respect to
$\pi N\to \Lambda K$ - we shall discuss here in detail only the case
of $<s\bar s>$ dependence on $<n_{coll}>$. 

Basic assumptions of our application of multiple nucleon- nucleon 
collision model to  $<s\bar s>$ production in
pA and AB collisions can be stated as follows:

 a) $<s\bar s>$ in pA and AB interactions is given  
 as the sum of $s\bar s$ pairs produced in
individual nucleon-nucleon (nn) collisions, with the number of nn collisions
calculated by the Glauber model. This
assumption is equivalent to the statement that the 
contribution to $<s\bar s>$ due to
"intrinsic strangeness" which appears in fragmentation of nucleon
remnants after the last interaction of a nucleon 
is much smaller than the contribution due to 
$s\bar s$ production in nucleon- nucleon interactions. 
In this sense total strangeness production is  assumed to be 
similar to hard or semi- hard processes
 and the number of strange quark pairs 
   produced in AB interaction is in the first
approximation proportional to the number of nucleon-nucleon collisions.

  b) The proportionality is assumed to be modified by the
attenuation of strangeness production with increasing number of previous 
interactions of the two nucleons participating in a nucleon- nucleon
collision.

These assumptions require a few comments. Data \cite{r8,r9} on
$\Lambda $ production in pA interactions has been parametrized
\cite{r8,r9} as
\begin{equation}
 \sigma ^{\l}_{pA}=A^{\alpha}\sigma_{pp}^{\l}
\label{eq3}
\end{equation}
with $\alpha $ close to 1. In a picture of multiple collisions where
$s\bar s$ are produced in nucleon- nucleon interactions and the s-quark
recombines with valence u- and d- quarks to form $\l$
 such an $A^{\alpha}$ dependence
is expected. 

By making the assumption a) we are in fact extending to $s\bar s$ the
description which is standard for production of $c\bar c$ and of
heavier quark pairs. $c\bar c$ pairs are assumed to be produced by
parton- parton, mostly gluon- gluon (gg) interactions. In the cms
frame of nucleon- nucleon collision in the CERN SPS energy region,
 momentum fractions of gluons producing $c\bar c$ pairs are about
x=0.15. Cross- sections for these energetic gluons are rather small
and the corresponding components of gluon structure functions are
only very little attenuated during the passage of a nucleon through
nuclear matter. Gluons responsible for $s\bar s$ production have 
lower values of x and we expect that the attenuation of the
corresponding part of structure function would be larger than in
the case of $c\bar c$ production. Because of that we have introduced
also the assumption b). We shall determine the amount of attenuation 
by comparing the model with the data on p- and light ion- induced
nuclear collisions. In order to estimate the errors of the extrapolation
to the Pb+Pb case we shall study two parametrizations of the attenuation.

  In the former version of our model (referred to as Model 1) we shall
  calculate the attenuation via the decreasing cms energy squared of
  nucleon- nucleon collision. In each collision the nucleon loses a part
  of its momentum and the energy squared ($s_{ij}$) for the collision
  which is $i$-th for one nucleon and $j$-th for the other one is decreasing
  with increasing $i$ and $j$.
  The dependence of $s_{ij}$ on $i,j$ is parametrized as follows
$$ y_{ij}=y_{beam}-[(i-1)+(j-1)]\Delta y_{lost} $$
\begin{equation}
s_{ij}/GeV^2=2+e^{y_{ij}}+e^{-y_{ij}}
\label{eq4}
\end{equation}
 Values of $<s\bar s>$ as a function
  of $s_{ij}$ are taken from the data on total strangeness production
  in pp collisions.

 The model gives a useful parametrization of
  strangeness production but we cannot imagine that the nucleon in
  individual nucleon- nucleon collisions is slowed down as a whole
  and that its parton distribution is "rescaled" after every collision 
  to the new value of nucleon's momentum. Such a picture would
  contradict \cite{r17}  data on production of Drell- Yan pairs which
  show that momenta of valence (and in general faster) partons are
  not modified during the passage of nucleon through nuclear matter.

  In the latter version of our model (referred to as Model 2) 
  we shall simply assume that  $s\bar s$ production in $i$-th
  collision of one nucleon and $j$-th collision of the other one
   is proportional to
  \begin{equation}
  (1-\beta)^{i-1}(1-\beta)^{j-1}
  \label{eq5}
  \end{equation}
  Physics behind this assumption is rather simple. We imagine that
  $s\bar s$ production is due to parton- parton (mostly gluon- gluon)
  interactions with gluon momenta of the order of 0.3- 1.0 GeV/c in
  the nucleon- nucleon cms. In each nucleon- nucleon collision the
  number of gluons in a nucleon is depleted \cite{r18} by a fraction
  of $(1-\beta)$. This leads to attenuation of $s\bar s$ production
  in subsequent collisions.

In this note we shall not discuss the question of the production
mechanism of pions, which is not well understood at present, since
the data can be described both by the wounded nucleon model \cite{r10,r19}
and within the scheme of multiple nucleon- nucleon interactions \cite{r20}.

  We shall now present some details on the model of multiple nucleon-
   nucleon collisions we shall use in our computations.

 A simple version of the multiple collision model
  has been introduced by C.Y. Wong \cite{r20,r21,r22}.
It has been used to describe successfully 
total pA and AB cross sections \cite{r20}, transverse energy distributions
\cite{r23}, rapidity and $p_T$ distributions \cite{r20,r24} and
nuclear stopping power \cite{r25}.

The assumptions of the model are   

a) proton - nucleus and nucleus-nucleus
collisions can be decomposed as a collection of nucleon-nucleon collisions.

b) positions of nucleons in nuclei are uniformly distributed within the
sphere of radius $R_A=1.2A^{1/3} fm$
(we check that changing uniform nucleon density to 
Wood-Saxon type density changes our results by less than 10\%)

c) only binary collisions of target and projectile nucleons are considered
(no rescattering)

d) the number of nucleon-nucleon collisions is governed by inelastic cross
section of 32$fm^2$.

The main virtue of the model is its simplicity. It is
probably the simplest possible
representation of nucleus-nucleus interactions in terms of
nucleon-nucleon collisions.
 
In the former version of the parametrization of attenuation we use
Eq.(4). The s-dependence of $s\bar s$ production in nucleon- nucleon
collision is parametrized as
\begin{equation}
<{s\bar s}>  = 0.38*ln(\sqrt s) - 0.498
\label{eq6}
\end{equation}
where $\langle {s\bar s} \rangle$ is the average number
of $s\bar s$ pairs per event as defined in (1). 
$s=(p_1+p_2)^2/GeV^2$ where
$p_1$ and $p_2$ are four momenta of colliding nucleons.

 Eq.(6) is taken to reproduce our $<s\bar s>$ value
at 200 GeV/c and $<s\bar s>=0.11\pm0.02$ from \cite{r16}.
 
The degradation of 'nucleon' momentum or better the degradation
of that component of nucleon which is responsible for strangeness
production is given by $\Delta y_{lost}$ in Eq.(4). Comparison of data
with Monte Carlo calculation based on the model of multiple nucleon-
nucleon- collisions gives
$$ \Delta y_{lost}\approx 0.35  $$
Results of this version of the model are compared with data
 in Table I.

 Note that the value of $\Delta y_{lost}$ is smaller than that obtained
 from the data on proton rapidity distributions (nuclear stopping
 power) or from pion rapidity distributions described within the
 Glauber model. We shall discuss this rather important point in the
 next Section.

 In the second version of our model we  describe the attenuation
of the production of $s\bar s$ by Eq.(5) taking $\beta $ as a free
parameter. Proton- and light ion- induced data lead to
$$ \beta \approx 0.12  $$

 A comparison of results of calculations with the data is
given in Fig.3 and in Table I. 
As can be seen from Table I the extrapolation to the case of Pb+Pb 
interactions by using the two models leads to $<s\bar s>$= 330 in
Model 2 and to $<s\bar s>$= 270 in Model 1. Taking the difference
between the two models as an indication of the magnitude of the
error of extrapolation,  the measurement of $<s\bar s>$
 outside of the interval  $<s\bar s>$ = 300$\pm $30 in Pb+Pb 
 interactions would
 provide an evidence of the onset of a new dynamical regime.

 For consistency check we present in Table I
also the numbers of participants and collisions.

\section{Comments and conclusions}
\label{comments}
The attenuated linear dependence of $<s\bar s>$ and of $<\Lambda>$, $<K>$
and $<\bar \Lambda>$ on $<n_{coll}>$ is rather difficult to understand,
since it is similar to that of hard 
or semi- hard processes, like $J/\Psi$
and Drell- Yan pair production. The inclusive cross- section
 for e.g. $<\Lambda>$ production in pA collisions is given as
\begin{equation}
\sigma^{\Lambda}_{pA}=<\Lambda>_{pA}\sigma _{pA}
\label{eq7}
\end{equation}

With $<\Lambda>_{pA}\sim <n_{coll}>\sim A^{1/3}$ 
and $\sigma_{pA}\sim A^{2/3}$ we
obtain $\sigma_{pA}^{\Lambda}\sim A^{\alpha}$ 
with $\alpha$ close to 1, which
is just the experimental result in Eq.(3).
Approximate proportionality of 
 $<\Lambda>$ to the number of nucleon- nucleon collisions is thus
an experimental fact for pA interactions. Our results in Figs.1-3 and
in Table I just show that this approximate proportionality can be
extended also to the case of AB interactions.
  The proportionality indicates that in AB collisions final state
interactions do not strongly modify particle fractions. This is less
surprising for $<s\bar s>$ since it gets changed only by reactions like
$\pi N\to K\Lambda$ than for $<\Lambda>, <\bar\Lambda>$ and $<K^0_s>$
which are influenced also by processes like $N\bar\Lambda\to K\bar NN$ and
$\bar KN\to \Lambda\pi$. The failure of this simple picture for 
Pb+Pb interactions would give evidence for the onset of a new dynamical 
regime  which would lead, via final state interactions, or QGP
 to  $<s\bar s>$, $<\Lambda>$, $<\bar\Lambda>$ and $<K>$ very different from
the expectations based on the simple model described above. Such an
evidence would complete the one already given by the data of NA-50
Collaboration on $J/\Psi$ suppression \cite{r26}.

In analyzing data on $<s\bar s>$ production we have used two models
referred to as Model 1 and Model 2.

In the latter the attenuation is of $s\bar{s}$ production in
subsequent nucleon- nucleon collisions has been parametrized by Eq.(5)
with a rather simple interpretation described below Eq.(5). For a truly
hard process, like e.g. Drell- Yan pair production, one would obtain
$\beta $=0. The value we have found from data on $<s\bar s>$,
$\beta \approx $0.12 indicates that the $s\bar{s}$ production for 
light- ion ( up to Sulphur) induced nuclear interactions may be viewed
as a hard process with the attenuation factor given by Eq.(5).

The former model, specified by Eqs.(4) and (6) parametrizes $s\bar{s}$
production in a way frequently used for describing soft processes,
like pion production, in the Glauber model. A hard process, like 
Drell- Yan production, can be described also in this scheme by
chosing $\Delta y_{lost}$=0. The value of the  
parameter $\Delta y_{lost}$ found in our analysis is significantly
lower than the value usually obtained in analyses of data on pion
production or nucleon rapidity distributions, which are typically
in the interval 0.7 - 1.0. A particular case will be mentioned shortly.
Lower value of $\Delta y_{lost}$ found in our analysis of $s\bar{s}$
production indicates that $s\bar{s}$ pair are originated by a mechanism
which is harder than that responsible for 
pion production.

In a recent study Jeon and Kapusta \cite{r27} 
(see also the preceding work \cite {r28,r29} ) have used a Glauber model
approach to describe nucleus- nucleus interactions as a sequence of
binary nucleon- nucleon collisions. They have obtained agreement with 
data on proton and negative hadron rapidity and transverse momentum
spectra but failed to reproduce data on production of strange particles
in S+S interactions.  They have obtained only 80\% of
observed charged kaons, 50\% of observed neutral kaons and $\Lambda$'s
and 10\% of observed $\al$'s. The difference between their results and
ours can be traced back to the value of the parameter $\Delta y_{lost}$.
The value of $\Delta y_{lost}$ used by Jeon and Kapusta \cite{r27}
is much larger than ours. In particular for proton- nucleus
interactions the value of rapidity loss of the incoming proton in each
 nucleon- nucleon collision as calculated from the Eq.(13)
in Ref.\cite{r27} is $0.6$ units of rapidity.

The comparison of these values with $\Delta y_{lost}$=0.35 found in our
analysis of $s\bar{s}$ production indicates again that $s\bar{s}$ pairs
are produced by a mechanism which is harder than pion production and
nucleon deceleration.

The discrepancy between  values of $\Delta y_{lost}$ corresponding
to  pion and $s\bar{s}$ production indicates that different  
components of the nucleon structure function are attenuated
differently during the passage of a nucleon through nuclear matter.

Finally let us compare our picture of $s\bar{s}$
production with  other models.
An  approach based on the wounded nucleon model
has been presented by Kacperski \cite{r30}. The author
considers three types of nucleon-nucleon collisions: fresh-fresh,
fresh-wounded and wounded-wounded.  Strange particle production
 in these three types of collisions is different. This allows to fit
data on strangeness production. In our model
 the production of total strangeness in a particular nucleon- nucleon
collision depends on the number of preceding collisions of both nucleons 
and not only on whether the nucleon is wounded or not.

Another phenomenologically successful description of data
 motivated by the additive quark
model has been presented by Kadija et al. \cite{r31}.
 The number of produced $\l$'s and $\al$'s is assumed to be
 proportional to the number of collisions of constituent
 quarks. Each nucleon can provide at most three constituent quarks.
In order to describe data one has to introduce rescattering in the final
 state. 

 Our picture is close to that of Kadija et al. \cite{r31}
in assuming that strangeness is produced by a mechanism which is
harder than presumably rather soft fragmentation process.

 Geist and Kachelhoffer \cite{r32} study production of strange
 particles in pA collisions and parametrize  
 the A-dependence of  production cross section by
  $$A^{0.8-0.75x+0.45x^3/|x|}$$
 where $x$ is the momentum fraction 
carried by a strange particle.
This leads to $A^{0.5}$ dependence in the forward region,
to $A^{0.8}$ in the central region and to $A^{1.1}$ in the backward region.
This parametrization is more detailed than ours but when integrated
over x 
 such a dependence on $A$ seems to be compatible with our 
version of the multiple collision model with
 attenuated $<s\bar s>$ production. The x-distribution of final
state strange hadrons is influenced by the recombination process
and the strange quark picked up by two valence quarks may have a
value of x typical for the fragmentation region. Thus although $s\bar s$
pair is produced originally at small x, the $\Lambda $ and $\bar \Lambda$
carrying s and $\bar s$ may have quite different value of x. 

Capella et al. \cite{r33} has shown that rapidity distribution of $\l$
and $\al$ as observed in nucleus- nucleus collisions at the CERN SPS,
can be reproduced by the independent string model with diquark-
antidiquark pairs in the sea, flavour symmetric quark sea
 and with final state interactions of
secondary hadrons. These final state interactions are responsible 
for the increase of $\l$ production via $\pi N\to K\l$. The model of
Ref. \cite{r33} is based on a different picture of strangeness 
production than ours and a direct comparison is impossible. Let us note
only that the presence of final state interactions of the type
$\pi N\to \l K$ would bend the curves in Fig.1 upward for higher
values of AB. 
 
Models based on string fusion \cite{r3,r4} combined with
final state interactions cannot be directly compared with our simple 
picture of attenuated gluon- gluon interactions.

Before concluding let us make a comment concerning a
possible origin of the attenuated linear dependence of $<s\bar s>$
on $<n_{coll}>$. In the Geiger partonic cascade model \cite{r35} of heavy
ion collisions in the CERN- SPS energy region
the perturbative QCD parton production
 provides an important contribution
to the rapidity density of final state
 particles in the central region \cite{r36}.
Being harder than fragmentation the partonic cascade may lead to
an increased $<s\bar s>$ production.  

 {\bf To conclude:} we have shown that a simple model based on two
 main assumptions namely 

a)  nucleus-nucleus interactions of protons and lighter ions
   can be considered as  superpositions of incoherent
   nucleon- nucleon collisions, the total strangeness being given as
   a sum of contributions of nucleon- nucleon interactions

 b) cross- section for production of 
    strange quarks pairs  in nucleon- nucleon collision is
    attenuated with the increasing number of preceding collisions.
    
\medskip
can consistently describe, within 1-2 std. errors,  data on total
strangeness  production in proton-nucleus and nucleus-nucleus
collisions up to S+Ag. The discrepancy between the value of 
$\Delta y_{lost}$ found from $s\bar{s}$ production and that from
pion rapidity distributions indicates that different components of
nucleon structure function are attenuated differently during the
passage of nucleon through nuclear matter.

The extrapolation of the model to the case of Pb+Pb interactions,
together with the estimate of the extrapolation uncertainity based
on the difference between the two versions of the attenuation, 
provides an estimate of total strangeness production in Pb+Pb
interactions in a situation where no new dynamical regime sets on:
$$ <s\bar s>_{Pb+Pb} \approx 300 \pm 30 $$
Data larger than this estimate would give an evidence of the presence
of a new dynamical regime, complementing the one provided by data on
$J/\Psi$ suppression \cite{r26}.

{\bf Acknowledgements} We are indebted to
 W.Geist, R.C.Hwa, I.Kr\'alik and  M. Moj\v zi\v s  
 for numerous useful discussions
 and to R.C.Hwa, N.Pi\v{s}\'utov\'a and P.Z\'avada
for collaboration at the early stages of this work. One of the authors
 (J.P.) would like to thank to G.Roche and B.Michel for the hospitality
at the Laboratoire de Physique Corpusculaire at the Blaise Pascal
University at Clermont- Ferrand.  

Last but not the least we are indebted to the unknown referee for
valuable comments and recommendations.

\medskip

{\bf Table I}
The dependence of the number of the strange quark
pairs per event $<s\bar s>$
on the number of nucleon-nucleon collisions in pMg,pS,pAr,pAg,
pXe,pAu,SS,SAg collisions calculated in model 1 (column 6)
and model 2 (column 7). For comparison data are presented
in column 5. In column 3 we give the number of nucleons participating
at least in one collision (Part.) and the next column gives the
number of nucleon- nucleon collisions (Coll.).

\bigskip

\begin{center}

\begin{tabular}{|c|c|c|c|c|c|c|c|}  \hline
 &$p_{LAB}$ & Part.& Coll.&$\langle {s\bar s}
\rangle$ DATA & $\langle {s\bar s}\rangle$
MOD 1& $\langle {s\bar s}\rangle$
MOD 2 & ref.  \\ \hline
   &    &     &      &     &       &                  &   \\     
pp & 200 GeV/c &2   & 1& 0.63$\pm$0.08 &- & &   \\  \hline
   &           &    &  &               &   &  &               \\
pMg& 200 GeV/c & 3.3&2.3& 0.9$\pm$0.3& 1.2 &1.2 & \cite{r9}     \\ \hline
   &           &    &   &              &      &          &   \\   
pS & 200GeV/c  &3.4&2.4& 1.40$\pm$0.23&1.3 & 1.3& \cite{r7}  \\  \hline
    &    &       &       &   & &     & \\
pAr &200GeV/c & 3.5&2.5& 0.84$\pm$ 0.14& 1.4 &1.4 &\cite{r8}   \\ \hline
    &    &       &       &   & &   & \\
pAg& 200GeV/c &4.4 & 3.4& 1.9$\pm$0.3 & 1.7 & 1.7 & \cite{r9} \\  \hline
    &          &                 &         &        &        &&             \\
pXe& 200GeV/c & 4.6& 3.6& 1.3$\pm$ 0.3& 1.8& 1.8 & \cite{r8} \\ \hline
    &          &                 &         &          &      &&             \\
pAu & 200 GeV/c & 5. &4.&1.7$\pm$0.30& 1.9& 1.9 & \cite{r9}   \\  \hline      
    &          &    &   &         &    &     &    \\
SS  & 200GeV/c & 56 &72 & 38.0$\pm$ 5.4 & 35 & 35 &\cite{r7} \\ \hline
    &          &    &   &         &    & &          \\
SAg &200GeV/c & 89 &135& 55.5$\pm$ 6.5& 58 & 60& \cite{r7}   \\ \hline
    &          &    &   &         &    & &          \\
PbPb &160GeV/c & 404&905& -  & 270  & 330 & -    \\ \hline
\end{tabular}
\end{center}

\vfill
\eject
{\Large \bf Figure Captions}            

\medskip

{\bf Fig.1}

The dependence of $<\l>, <\al>$ and $<K^0_s>$ on the mean number 
of nucleon- nucleon collisions $<n_{coll}>$ for pp, pA and AB  
interactions.
The data are from Refs. quoted in Table I.

\medskip
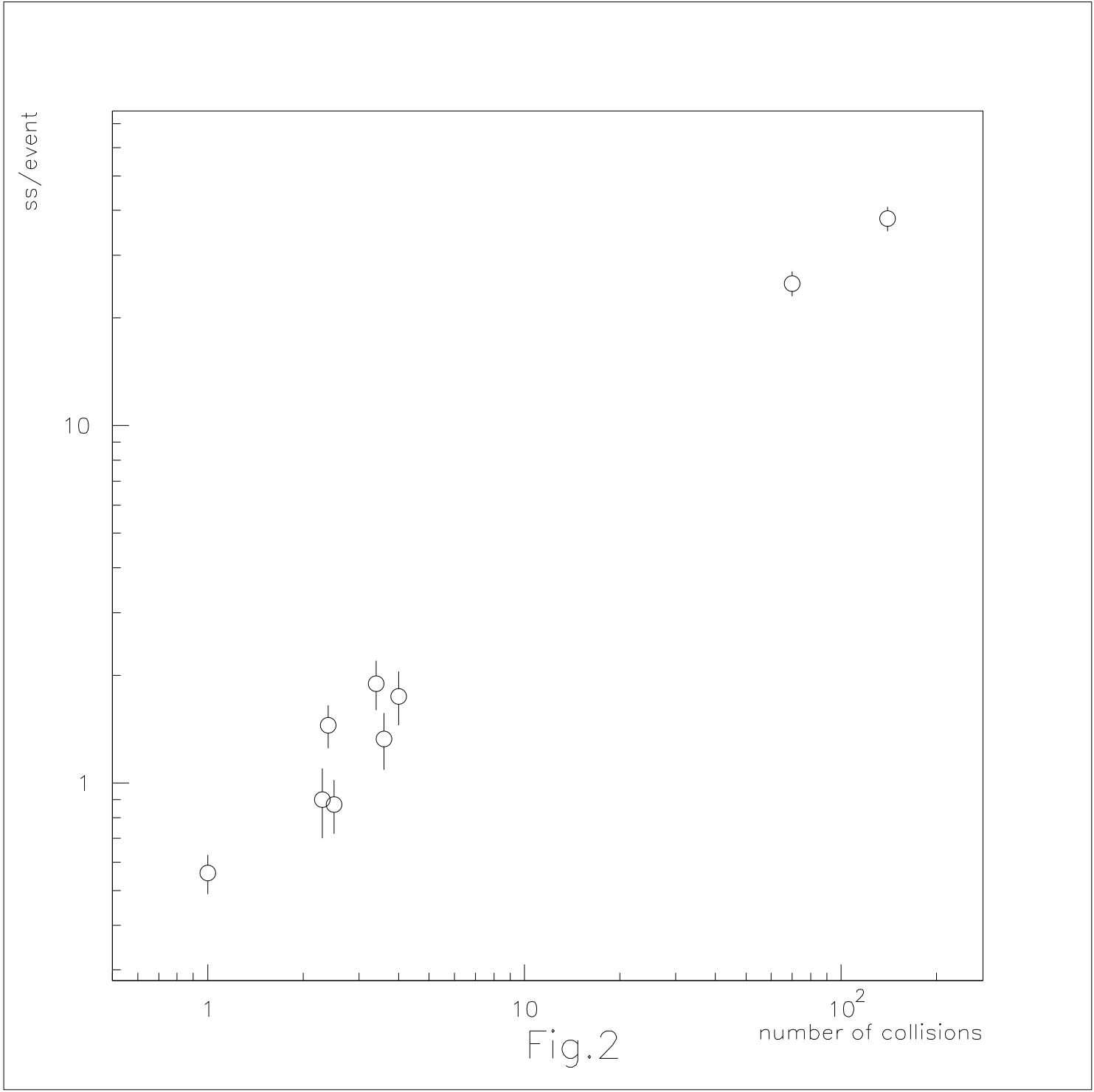
{\bf Fig.2}

The dependence of the mean number of $s\bar s$ pairs on the mean
number of nucleon- nucleon collisions $<n_{coll}>$ for pp, pA and 
AB collisions. Data are taken from Refs. given in Table I.

\medskip
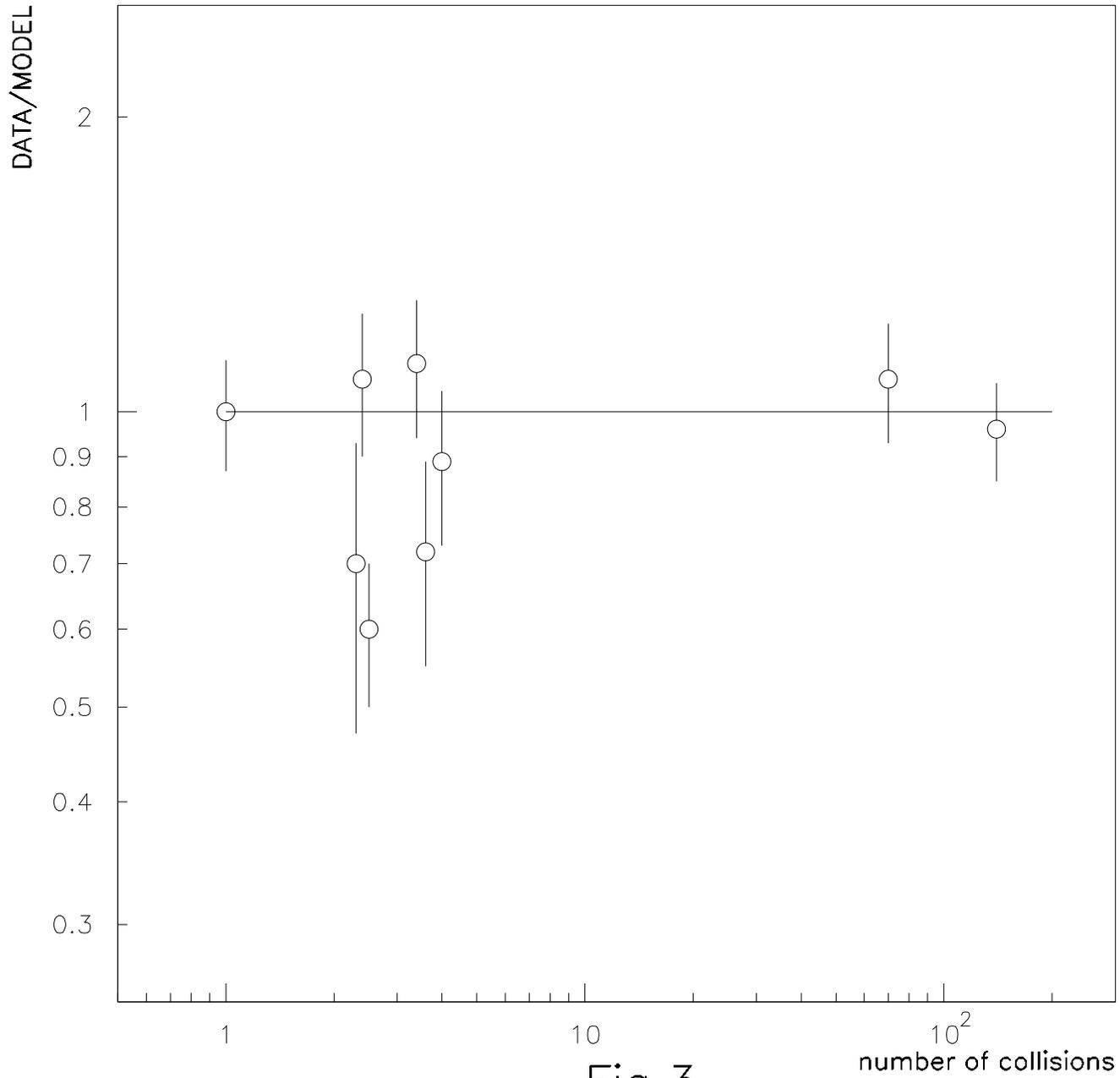
{\bf Fig.3}

The ratio of the data on $<s\bar s>$ in pp, pA and AB interactions
and results of our two models. Within the values of $<n_{coll}>$ shown 
the models give almost identical results.

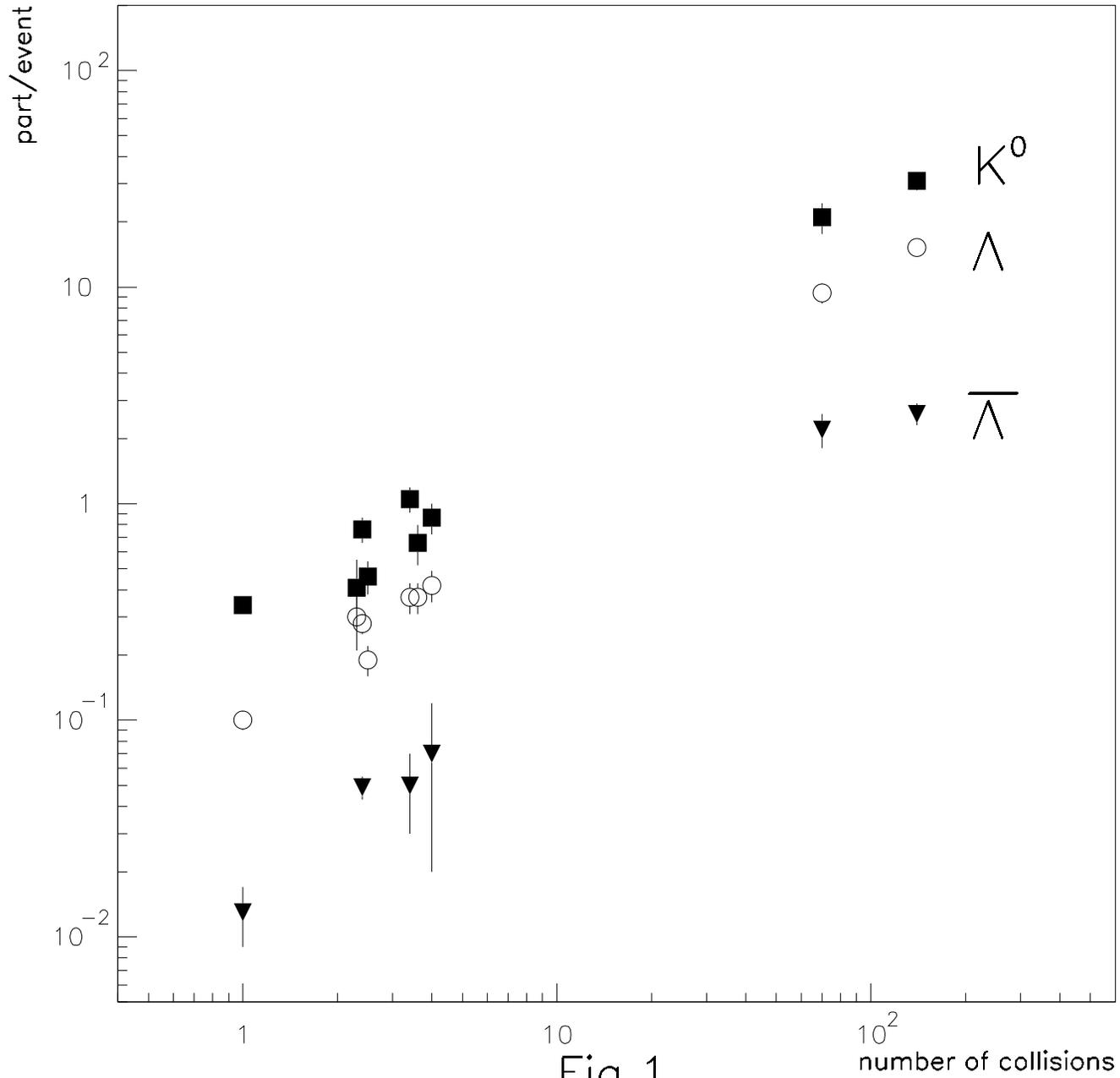
\end{document}